**Fluorescent diamond microparticles doped glass fiber for magnetic field sensing**

*Dongbi Bai, Minh Hoa Huynh, David A. Simpson, Philipp Reineck, Shahraam A. Vahid, Andrew D. Greentree, Scott Foster, Brant C. Gibson, Heike Ebendorff-Heidepriem\**

Dr. D. Bai
School of Science
RMIT University, Melbourne, VIC 3001, Australia
E-mail: dongbi.bai@rmit.edu.au
M. H. Huynh, Prof. H. Ebendorff-Heidepriem
E-mail: heike.ebendorff@adelaide.edu.au
Institute for Photonics and Advanced Sensing, School of Physical Sciences
The University of Adelaide, Adelaide, SA 5005, Australia
Dr. D. A. Simpson
School of Physics
The University of Melbourne, VIC 3010, Australia
A/Prof. S. A. Vahid
School of Engineering
University of South Australia, Adelaide, SA 5095, Australia
Dr. S. Foster
Defence Science and Technology Group, Edinburgh, SA 5111, Australia
Dr. P. Reineck, Prof. A. D. Greentree, Prof. B. C. Gibson
ARC Center of Excellence for Nanoscale BioPhotonics
RMIT University, Melbourne, VIC 3001, Australia
E-mail: brant.gibson@rmit.edu.au

Keywords: Fiber design and fabrication, fiber sensors, magneto-optic systems, fluorescent materials

Diamond containing the negatively charged nitrogen-vacancy (NV) center is emerging as a significant new system for magnetometry. However, most NV sensors require microscopes to collect the fluorescence signals and are therefore limited to laboratory settings. By incorporating micron-scale diamond particles at an annular interface within the cross section of a silicate glass fiber, a high-sensitivity and robust fiber platform for magnetic field sensing is demonstrated here. The fluorescence and spin properties of NV centers embedded in the diamond crystals are well preserved during the fiber drawing process, leading to enhanced continuous-wave diamond-magnetometry in fiber-transmitted sensing configurations. The interface doping of diamond particles also leads to reduced fiber propagation loss and benefits the guidance of NV-fluorescence in the hybrid fiber. Using the diamond-fiber system, magnetic field readout through 50 cm of fiber is achieved. This study paves the way for novel fiber-based diamond sensors for field-deployable quantum metrology applications.

# 1. Introduction

Nitrogen-vacancy (NV) colour centers in diamond have been extensively studied for precise magnetic field sensing, [1-4] quantum information technology, [5] as well as bio-medical thermometry and imaging. [6,7] As a sensor, the NV center can be optically addressed and manipulated remotely. Conventional NV magnetometer systems use either confocal [8,9] or widefield [10] collection of the fluorescence to provide rich information of their local (potentially nanoscale) environment. Nevertheless, the sophistication and lack of portability of such high-end microscopes has led to interest in integrating diamond NV centers with glass optical fibers. To date, the most commonly used approach to realize such a hybrid device relies on growing or placing diamond materials external to the fiber, [11-14] which enables NV emitters to locate either on the endface of the fiber [15-17] or tapered fiber regions. [18-20] Diamond nanoparticles have been combined with tellurite (Te) glass to create hybrid diamond-glass fibers,[21] and Mayani *et al.* demonstrated distributed diamond-fiber magnetometry by periodically spacing photodiode chips along a polycarbonate cladding to detect the diamond-microfluidic systems configured in a hollow-core silica fiber. In combination with lock-in amplification, they reported DC magnetic field sensitivity of ~63 nT/√Hz.[22]

Previously, we have employed a volume doping technique to integrate fluorescent diamond nanoparticles into the glass matrix, where the nanodiamonds (ND) are dispersed into molten glass to create a ND-doped glass billet, which is subsequently processed using preform extrusion and fiber drawing to obtain an optical fiber with NDs distributed inside the whole fiber volume. [23, 24] The volume doping approach has received remarkable progress in maintaining the single-photon emission [21, 24] and optically detected magnetic resonance (ODMR) [25] properties of NV centers in fluorescent NDs embedded in fiber. However, the volume doping technique is challenging in terms of oxidization and dissolving of diamond nanoparticles during the doping step [24, 26], as well as achieving efficient excitation and collection of NV-fluorescence via the fiber due to the absorption and scattering from diamond particles randomly distributed over the whole fiber cross-section.

Here we report a diamond-doped optical fiber made from lead-silicate glass using micron-sized NV-containing diamond crystals and a new doping method, which confines the diamond particles at an annular interface within the fiber cross section. The resultant fiber presents low loss and spatial confinement of NV-fluorescence to the central region of the fiber geometry under laser illumination, which are favourable for fluorescence signal collection along the fiber. We performed magnetic field sensing in ambient environmental conditions via three different fluorescence excitation and collection architectures relative to the fiber axis. The magnetic field sensitivity of the diamond-doped fiber is considerably enhanced due to the utilization of microdiamonds and the configuration of the diamond particles on the ring-shaped interface. We detected bright photoluminescence from NV centers at the fiber output end, which leads to a fiber-based diamond-magnetometer with a sensing length up to 50 cm. The hybrid fiber also allows for the convenience of operation without easy breakage, and promises the compatibility of magnetically sensitive fiber with other fiber systems. Our results indicate that the microdiamond-doped lead-silicate glass optical fiber can be applied in versatile fiber platforms for remote magnetometry.

**2. Results and Discussion**

**2.1. Fiber fabrication and characterization**

We selected commercial F2 lead-silicate glass (Schott Glass Co.) [27] as the fiber material since this glass has a high mechanical stability and can be easily drawn into different structures without crystallization. [28, 29] Diamond particles with average diameter of ~1 μm were used as dopants (see Experimental Section) since such large particles offer higher number of NV centers per particle and improved quality of the diamond host material compared to nanoscale diamond particles, which are advantageous for enhancing the magnetic field sensitivity. [2, 30] **Figure 1**a illustrates the fiber fabrication steps, which followed a cane-in-tube approach. [28, 31] Firstly, a rod with outer diameter (OD) of ~11 mm and a tube with OD of ~9 mm and inner diameter (ID) of ~0.9 mm were made from ~30 mm diameter F2 glass billets using the extrusion technique. [31, 32] Secondly, the rod was scaled down in size to OD of ~0.6 mm using a drawing tower. Thirdly, the surface of the cane was coated with diamond particles by dipping the cane into a diamond solution by using a dip coater (see Experimental Section), and subsequently dried in air. Finally, the diamond-coated cane was inserted into the tube, and this preform assembly was drawn down to fiber using a drawing tower. Reduced pressure was applied during fiber drawing to close the gap between the cane and tube. From a single cane-in-tube assembly, over 100 m length of fiber were drawn. The cane (tube) part of the preform assembly leads to the inner (outer) region of the fiber. The diamond particles are distributed at the ring-shaped inner/outer interface within the fiber cross-section. Hereafter, we refer the fiber fabricated from F2 cane and F2 tube as F2/F2 fiber.

Figures 1b and c show optical images of the diamond-coated F2 cane and the cross-section of the diamond-doped F2/F2 fiber, respectively. We took fiber samples from different bands of the fiber, and measured the fiber OD and the diameter of the inner/outer fiber interface to be ~130 µm and ~9 µm, respectively. We observed tiny air holes at the fiber inner/outer interface, as shown in Figure 1d, from all bands of the drawn down F2/F2 fiber. To determine whether the air holes are caused by the embedded diamond particles, we fabricated an undoped F2/F2 fiber by using the same cane-in-tube preform parameters and fiber drawing conditions but without coating the cane. Similar interface hole features were shown in all sections of the undoped F2/F2 fiber. We note that the same effect of interface holes was observed for a step-index Te fiber, which was also made by the cane-in-tube technique with the cane and tube glasses having similar glass viscosity at the drawing temperature.[33]

Therefore, we attribute the interface holes in the diamond-doped F2/F2 fiber to the fact that the F2 glass cane and F2 glass tube experienced the high and identical glass viscosity of ~$10^6$ dPa.s during the fiber drawing process, [32] which causes incomplete fusion of the F2 glass cane with the F2 glass tube. For further comparisons, we also fabricated an unstructured undoped F2 fiber without any interface, where the extruded rod preform was directly drawn down to a fiber with ~160 μm diameter.

Figure 2 shows the optical loss spectra for the diamond-doped F2/F2 fiber, undoped F2/F2 fiber and undoped unstructured F2 fiber (see Experimental Section). The measured propagation loss at the excitation wavelength of 532 nm for NV-fluorescence is higher for the diamond-doped F2/F2 fiber (~4.6 dB/m) compared to both the undoped F2/F2 fiber (~2.4 dB/m) and the undoped unstructured F2 fiber (~2.1 dB/m). Likewise, the propagation loss at the NV-fluorescence emission wavelength range of 600-800 nm is higher for the diamond-doped F2/F2 fiber (~4.0 dB/m) compared to the undoped fibers (~2.0 dB/m). The similar loss for the two undoped fibers indicates that the air holes at the inner/outer interface for the diamond-doped F2/F2 fiber induce only a small increase in fiber loss relative to the F2 fiber without interface holes. Therefore, the higher loss of the diamond-doped F2/F2 fiber relative to the undoped fibers is attributed to the micron-sized diamond particles at the inner/outer interface of the fiber.

## 2.2. Fluorescence and ODMR measurements

Three fluorescence excitation and emission collection schemes were implemented to investigate the magnetic field sensing behaviour of our diamond-doped F2/F2 fiber (see Experimental Section). These configurations are shown in Figure 3a and were side excitation (port A) with confocal collection (side/side scheme), side excitation (port A) with fiber endface (port B) collection (side/end scheme), and excitation from the proximal end (port C) with collection at the distal end (port B) of the fiber (end/end scheme). The end/end scheme corresponds to the architecture that we envisaged for remote magnetometry. For all three schemes, the data was experimentally obtained from the same piece of fiber sample. The fiber length from port A to B and port C to B was measured to be 20 cm and 50 cm, respectively. Figure 3b shows the fluorescence mapping result obtained from port A using the side/side scheme. The fluorescence is photostable with no blinking and is observed to originate from emitting particles located at the interface region of the F2/F2 fiber. Figure 3c corresponds to the fluorescence mapping result recorded from port B by using the side/end scheme, in which the fluorescence emissions show a radial spreading through the fiber cross-section. We further examined fiber pieces taken from all bands of the drawn F2/F2 fiber in different lengths and applied laser excitation either from the side or proximal end of the fiber. The same cross-sectional fluorescence distribution was observed for every fiber sample. According to all the fluorescent images we experimentally obtained, we estimate the inner region of the fiber delivers ~17 times higher fluorescence intensity than the outermost region of the fiber. We attribute the spatial confinement of high-index diamond particles at the fiber inner/outer interface influences fluorescence guidance in the F2/F2 fiber. Specifically, due to the four allowed NV orientations within a single diamond microcrystal, the diamond particles embedded in fiber fluoresce approximately isotropically under laser excitation. The photons emitted towards the fiber inner region have a higher probability to scatter off other diamond particles, and thus be gathered in the central region of the fiber.

In Figure 4a, we depict the fluorescence spectra recorded from emitting spots in the fluorescence images of the F2/F2 fiber for all three excitation/collection schemes. In all measured fluorescence spectra, the characteristic zero phonon line (ZPL) around 575 nm for neutral state NV centers ($NV^0$) and ZPL around 637 nm for negatively charged nitrogen-vacancy centers ($NV^-$) are clearly identified, confirming the NV centers in the diamond particles as the origin of the observed fluorescence. The components around 550 nm are attributed to the Raman peaks for F2 glass under CW 532 nm laser excitation, which become pronounced when the fiber was examined under longitudinal end/end excitation and collection scheme (green curve in Figure 4a). We note that because only the NV- centers possess the spin properties which are of interest in the diamond quantum sensing applications, it is essential to select an optimal combination of optical filters to minimize the background noise for the practical implementations of the diamond-doped F2/F2 fiber.

Diamond-magnetometry was investigated for the three excitation/collection schemes. Figure 4b shows the ODMR spectra recorded from the ensemble of NV centers in micron-size diamond particles embedded in the F2/F2 fiber, which were initialized using both the side/side and the side/end scheme. The excitation laser was launched from port A, and the NV- fluorescence as a function of scanning microwave (MW) frequency was simultaneously collected through the confocal setup via port A (black curve, side/side) and longitudinal transmission to the fiber end port B (blue curve, side/end). The ODMR dip at ~2870 MHz for NV centers is observed at zero magnetic field, which is further split by the Zeeman effect in the presence of an external magnetic field. The frequency transitions in the electron-spin resonance spectra thus provide the measurement of the vector components of the applied magnetic fields. [2,9] We observe that the spin-readout contrast for signals collected from the side/end scheme is generally lower than that from the conventional confocal side/side scheme. Furthermore, with increasing magnetic field strength, some vectorial information of the applied field is masked when the fluorescence emission is transmitted and detected along the fiber using the side/end scheme. We assume that this effect resulted from an interplay of fiber loss, light interference from the weakly excited diamond particles outside the focal point, as well as signal averaging in our ODMR acquisitions.

Driven by the light transmission capability of the fiber, we further detected the ODMR spectra via longitudinal end excitation/end collection fiber sensing scheme. As a proof-of-concept experiment, in the current end/end scheme, the excitation beam was transmitted along the whole fiber sample, the MW radiation was applied over the middle section of the fiber. Figure 4c shows the resultant ODMR spectra from the 50-cm fiber sample via end/end scheme, where the collected NV-fluorescence presents a read-out contrast of ~2.5%. The Zeeman splitting of the signal is distinguishable under external magnetic field strength of less than ~1 mT. We note that when stronger magnetic field is applied, the spin-dependent fluorescence transmitted to the fiber end becomes masked by the background luminescence from NV centers along the fiber which do not experience the MW excitation. Future study is required on reducing the fiber loss and optimizing the optical interrogation of controlled number of diamond particles along the fiber to realize high-fidelity magnetic field readout via remote end/end scheme.

We determine the sensitivity of our diamond-doped F2/F2 fiber to DC magnetic fields by $\eta_{dc} \approx \Delta\nu/(\tilde{\gamma}_e C \sqrt{I_{PL}})$, [2,3,9] where $\Delta\nu$ is the full width half maximum (FWHM) linewidth of the ODMR dip, $\tilde{\gamma}_e \approx$ 2.8 MHz/G is the electron gyromagnetic ratio, $C$ is the ODMR contrast, and $I_{PL}$ is the detected NV photoluminescence intensity under CW laser excitation. As a result, the F2/F2 fiber demonstrates a room-temperature DC magnetic field sensitivity of 350 nT/√Hz for localized NV-spin characterization (side/side scheme), 650 nT/√Hz for a fiber transmission length of 20 cm (side/end scheme), and ~3 µT/√Hz for longitudinally signal pumping and readout in a fiber length of 50 cm (end/end scheme).

We point out the F2/F2 fiber achieves considerably enhanced magnetic sensitivity for hybrid diamond-doped glass optical fibers. As a comparison, the previously reported ND-doped Te fiber showed 10 µT/√Hz sensitivity in a 5-cm side/end sensing scheme,[23] while in this study the F2/F2 fiber showed over one order of magnitude higher magnetic field sensitivity in a 20-cm side/end sensing scheme. The sensitivity improvement demonstrated by the F2/F2 fiber is attributed to several following effects. In Table 1, we further compare the relevant properties with the ND-doped Te fiber in reference 24 and 25.

*Diamond particle size:* The F2/F2 fiber was doped with micron-size diamond particles (~1 µm). For magnetometry with NV defects in diamond, the shot-noise limited DC magnetic field sensitivity scales with $\sqrt{N_{NV}}^{-1}$, where $N_{NV}$ is the number of NV centers used in the sensing process. [2] Neglecting any agglomeration effects, the ~1 µm diamond particles used for the F2/F2 fiber offers about four orders of magnitude more NV centers than nanoscale diamond particles (~45 nm, as used in the Te fiber) due to their larger volume, and therefore theoretically predicts a ~100 times enhancement in the sensitivity (without taking fiber loss into account). Furthermore, for diamond particles above 100 nm in size, more NV centers are in general in the $NV^-$ state compared to smaller particles. [30, 34] This is also important since only the $NV^-$ fluorescence is sensitive to magnetic fields.

*Doping geometry:* The F2/F2 fiber was made using the interface doping method, where the diamond particles are spatially confined to the annular interface between the inner and outer region. The interface embedding of diamond particles in fiber improves the optical transmission as evidenced by the lower loss of the doped F2/F2 fiber (~4 dB/m) relative to the doped Te fiber (~10 dB/m).[24] The comparison on fiber loss indicates that the spatial confinement of diamond particles compensates the larger scattering of the micron-scale diamonds relative to smaller nanodiamonds. The lower loss of the F2/F2 fiber is also beneficial for increasing the number of photons collected from the fiber endface. In addition, the placement of NV emitters at the fiber inner/outer interface gathers the NV emission in the central region of the fiber waveguide, which further improves the fluorescence detection efficiency in fiber-transmitted sensing scheme for enhanced ODMR sensitivity.

*Doping method:* For the interface doping technique used for the F2/F2 fiber, diamond particles are embedded in the glass during the fiber drawing step, in which the glass has an estimated viscosity of ~$10^6$ dPa.s (i.e. the glass is softened but not molten). This relatively high glass viscosity during the interface doping step reduces mass transport in the glass via convection and diffusion, which hinders the chemical reaction of the glass with the surface of the diamonds and/or dissolved gases in the glass. This property is essential to preserve the fluorescence emissions of NV centers in diamond particles embedded in the F2/F2 fiber, which leads to improved sensitivity in magnetic field sensing.

Moreover, our study proves the significance of the glass viscosity on the survival of nano/micro-crystals during the doping step. Comparing the interface doping for F2/F2 fiber and the volume doping for Te fiber, both the F2 and the Te glass were doped with diamond particles at a similar temperature of 600-650 °C using similar doping time of 10-20 min, [24] but with very different viscosity; softened glass of ~$10^6$ dPa.s for F2 glass and molten glass of ~$10^0$ dPa.s for Te glass. [35] In accordance with the high viscosity, no evidence for dissolution of diamond particles was found for the interface doping of F2 glass in this study, while strong diamond dissolution was found for the low-viscosity Te glass melt. [24] The critical effect of the glass viscosity at the diamond doping step is consistent with the observed increase on the dissolution of oxide nanocrystals in soft glass with increasing temperature.[36]

In addition, as shown in Table 1, the use of F2 glass in this work shows the advantage in the integration of the magnetically sensitive fiber to silica fibers. The relatively low refractive index of F2 glass (1.6) leads to better index-matching with silica glass (1.45), [27, 37] which promises reduced Fresnel reflection and thus higher light transmission. Moreover, F2 glass shows relatively high mechanical strength and low thermal expansion coefficient, [27] improving the mechanical robustness of diamond-doped fibers for practical applications.

## 3. Conclusions

In summary, we have developed the design and manufacture of intrinsic magnetically sensitive optical fiber by using lead-silicate glass, microdiamond particles and the interface embedding technique. The diamond-doped fiber is investigated as a robust sensor that drives the fiber-optic magnetometry with ensemble NV centers in diamond. We achieve a room-temperature DC magnetic sensitivity of ~350 nT/√Hz and ~650 nT/√Hz by characterizing the NV-fluorescence captured from fiber side and fiber output end, respectively. The fiber also facilitates ODMR readout via longitudinal laser initialization and signal collection along fiber axis, leading to a current magnetic sensitivity of ~3 µT/√Hz in a sensing length of 50 cm. The sensitivity of the fiber magneto-senor can be further enhanced by improving the purity of diamond particles with controlled nitrogen and carbon isotopes, [39-41] suppressing shot noise from the pump laser as well as magnetometry based on pulse quantum sensing and lock-in detection scheme. [42,43]

We also conclude that the interface embedding is provided as an effective approach to incorporate photonic emitters into fiber channels, bringing up new opportunities to generate optical fibers for different sensing applications. Furthermore, our results highlight the need of fabricating diamond-doped fibers with step-index core/clad structure, which yields potentially stronger coupling of NV-fluorescence to guided fiber modes and precise quantification of magnetic fields in longer sensing range. Our diamond-doped glass optical fiber can be expected to have significant impacts on quantum information science and persistent magnetic field monitoring applications.

## 4. Experimental Section

*Diamond Material*: Commercially available high-pressure high-temperature diamond particles (MSY 0.75-1.25, Microdiamond, Switzerland) were irradiated with 2 MeV electrons to a fluence of $1\times10^{18}$ cm$^{-2}$ followed by annealing (900 °C, 2 h) in argon to create NV centers. The particles were then oxidized in air (520 °C, 2h) to remove non-diamond carbon from the particle surface. Based on measurements on bulk diamond samples that were processed analogous to the particles used here,[39] we estimate the NV center concentration in diamond particle to be ~ 1 ppm.

*Cane coating*: The processed diamond powder was dispersed into ethanol to prepare a diamond solution with concentration of ~0.4 mg·mL$^{-1}$. The diamond solution was sonicated for over 30 minutes to reduce the agglomeration of diamond microparticles. The F2 glass cane was coated with diamond microparticles by using a dip coater set at a dipping/withdrawing speed of 200 mm/min. The remaining time in the diamond solution and the gap between each dip were both 30 s. We implemented 25 dips in the cane-coating process, which resulted in a relatively uniform distribution of diamond microparticles on the surface of the F2 cane (Figure 1b). We experimentally found that further dipping would wash off the diamond particles and, in some cases, increase the agglomeration of particles on the F2 cane.

*Fiber loss measurements:* We measured the fiber loss spectra in the range of 500 to 1600 nm by using the cutback method with a supercontinuum light source. The broadband light was collimated and focused into the fiber. Light transmitted out from the fiber was free-space coupled into an optical spectrum analyzer (AQ6315E, ANDO). We performed more than 4 cutbacks for each fiber, with 3 cleaves for each cutback length to minimize the measurement error.

*NV-Fluorescence Measurements*: A continuous-wave (CW) 532 nm laser (GEM532, Laser Quantum) was used as excitation light. We employed a 0.9 NA, 100× air objective to focus the excitation laser onto the fiber from fiber side (side/side and side/end schemes), and a 0.65 NA, 40× air objective to couple the excitation laser into the fiber from fiber proximal end (end/end scheme). For the side/side scheme, the same objective for excitation was used to collect the NV-fluorescence emission. For both the side/end and the end/end schemes, light transmitted out from the fiber was collimated by a lens with focal length of 30 mm for collection. A combination of 532 nm notch filter and a 600 nm longpass filter was used to separate the NV-fluorescence and the excitation laser. The filtered NV-fluorescence was collected by a multi-mode fiber (GIF625, Thorlabs), and then delivered to avalanche photodiodes (SPCM-AQRH-14, Excelitas Technologies) and a spectrometer (SpectraPro SP-2500, Princeton Instruments) for intensity and spectral measurement, respectively.

*ODMR Characterization*: We performed all ODMR measurements at room temperature by employing a CW 532 nm laser as pump source. Permanent magnets were used to vary the external magnetic fields. For both the side/side and side/end schemes, a microwave antenna was placed over the fiber close to the optical excitation. For end/end scheme, microwave antenna was transmitted across the middle part of the 50-cm fiber sample. A microwave generator (SMIQ03, Rohde & Schwarz) in combination with an amplifier (ZHL-16W-43+, Minicircuits) were used to deliver ~1 W of radio frequency power via the microwave antenna to drive the splitting of spin levels for NV centers. We experimentally detected a fluorescence emission rate of $10^7$ counts/s, $5.5\times10^6$ counts/s, $3.5\times10^6$ counts/s for ODMR signals obtained from the side/side, side/end, end/end sensing scheme, respectively. The three fluorescence excitation and collection schemes differ in their impact of fiber loss on the detected NV-fluorescence intensities as follows. For the side/side scheme, both the excitation and the emission of the NV-fluorescence experience negligible attenuation since they propagate only through a thin layer of glass (~50 μm) without diamond particles that would act as absorbing and scattering centers. For the side/end scheme, the fluorescence emission generated at port A is affected by the attenuation through 20 cm of fiber, whereas for the end/end scheme, both the excitation and the emission of the NV-fluorescence are affected by the attenuation through 50 cm of fiber.

**Acknowledgements**
This work was funded by the Defence Science and Technology Group under the Next Generation Technologies Fund (NGTF) program. Takeshi Ohshima and colleagues from QST, Takasaki, Japan are acknowledged for irradiating the diamond samples. D.B acknowledges Dr. Stephen Warren-Smith and Dr. Erik Schartner from the University of Adelaide for the fiber loss measurements and characterizations, Dr. Asma Khalid from RMIT University for the SEM imaging. A.D.G acknowledges the Australian Research Council for financial support (FT160100357). P.R. acknowledges funding through the RMIT Vice-Chancellor's Research Fellowship. This work was performed in part at the OptoFab node of the Australian National Fabrication Facility utilizing Commonwealth and SA State Government funding.

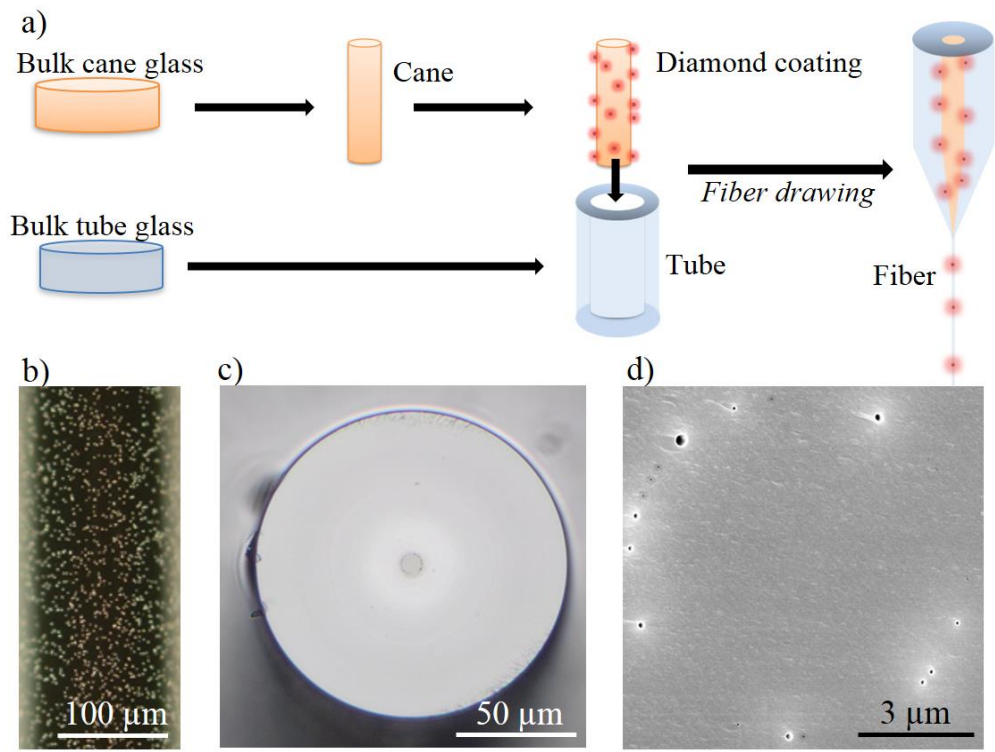

**Figure 1.** a) Illustration of the cane-in-tube fiber fabrication process. Two F2 glass billets were prepared. One billet was extruded into a tube. The other billet was extruded into a rod, which was drawn into a cane and coated with diamond particles. The coated cane was then inserted into the tube, and the cane-in-tube assembly was drawn down to the diamond-doped glass optical fiber. b) Optical microscope image of the F2 glass cane coated with 1-µm microdiamond particles. Diamond particles are presented as white dots against the black background. The image was taken before the cane was inserted into the tube for fiber drawing. c) Optical microscope image of the cross-section of the diamond-doped F2/F2 fiber. The fiber presents an outer diameter of ~130 µm. Diamond particles are doped at the interface between the inner and outer fiber regions. d) Scanning electron microscopy image of inner region of the diamond-doped F2/F2 fiber. The observed holes are due to the incomplete fusion of the F2 glass cane with the F2 glass tube in the fiber drawing process.

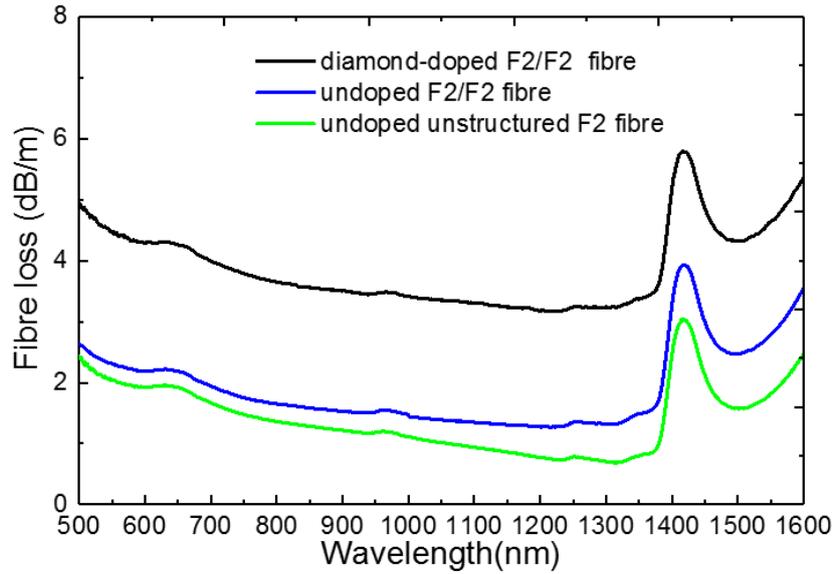

**Figure 2.** Propagation loss of diamond-doped F2/F2 fiber (black curve), undoped F2/F2 fiber (blue curve), and unstructured undoped F2 fiber (green curve) in wavelength range of 500-1600 nm. The relatively higher loss of diamond-doped F2/F2 fiber is mainly due to scatterings induced by the diamond particles. Comparison between the two undoped fibers indicates the interface holes have insignificant impacts on the fiber loss.

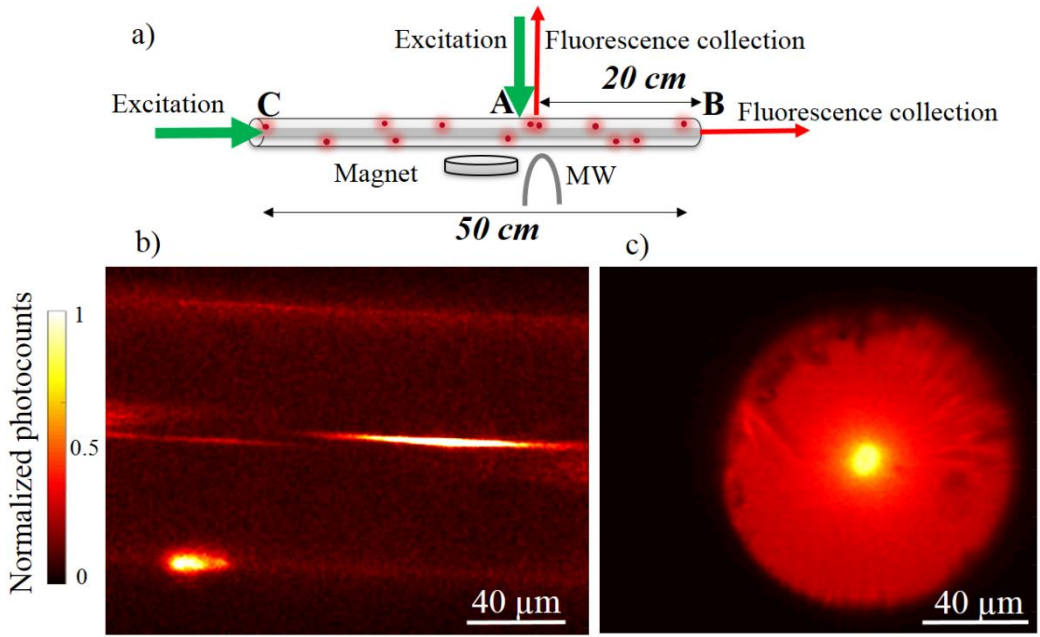

**Figure 3.** a) Schematic of fluorescence excitation and collection schemes for the diamond-doped F2/F2 fiber. In side/side scheme, the 532-nm continuous-wave laser was delivered from fiber side (port A) with confocal fluorescence collection at the same site. In side/end scheme, laser excitation was delivered at port A and fluorescence was collected at fiber endface (port B). In end/end scheme, laser excitation was launched from fiber proximal end (port C) and fluorescence was collected at the distal end (port B). For ODMR measurements, permanent magnets were used to vary the external magnetic fields. Microwave (MW) was provided by antenna placed close to the fiber. b) and c) Fluorescence mapping result obtained from b) the side of fiber using side/side scheme and c) the endface of fiber using side/end scheme. Bright spots in the central region of the fiber are fluorescent microdiamonds being confined at the fiber interface. The emitting spot on the outer edge of fiber in b) was caused

by random dust/oil contamination, which can be eliminated by improving the fiber sample preparation process.

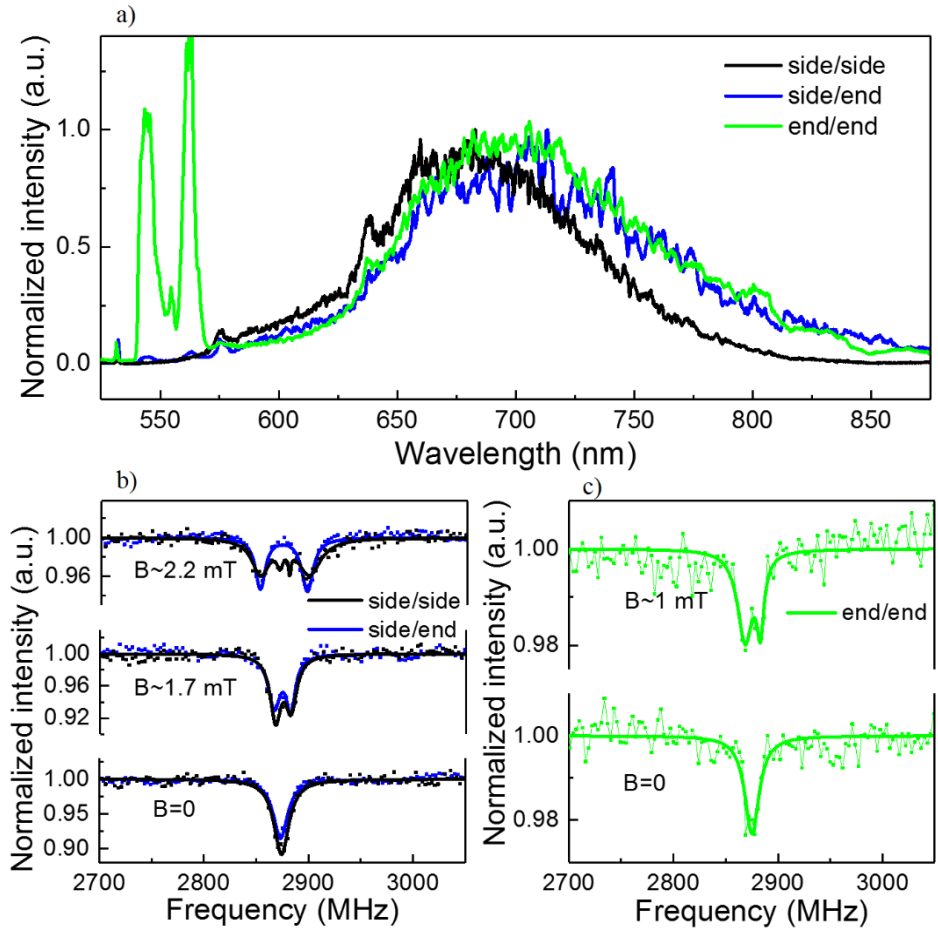

**Figure 4.** a) Optical spectrum in wavelength range of 525-875 nm recorded from fluorescent spots in different sensing schemes, confirming the NV centers in the diamond particles as the origin of the observed fluorescence in the F2/F2 fiber. The components around 550 nm from the end/end scheme are Raman peaks for the F2 glass under 532 nm laser excitation. b) ODMR spectra measured at increasing external magnetic field B. Black curve: fiber side excitation/side collection; blue curve: fiber side excitation/end collection. c) ODMR spectra recorded from longitudinal end/end sensing scheme. In b) and c), dots are experimental data, solid lines are Lorentzian fits.

**Table 1.** Comparison of the properties of the F2/F2 fiber of this study and the Te fiber in reference 24 and 25

| property | F2 | Te | F2 relative to Te | Refs |
|---|---|---|---|---|
| diamond particles size | 1 μm | 45 nm | larger size → higher NV density in excitation volume ☺ | 24 |
| doping geometry | interface | volume | high spatial confinement → lower loss of fluorescence ☺ | 24 |
| log viscosity (dPa.s) during the doping step[a] | 6 | 0 | higher viscosity → lower chemical attack of diamonds ☺ | 32,35 |
| refractive index contrast to silica[b] | 0.15 | 0.55 | smaller index contrast → higher coupling ☺ | 24,27, 37 |
| mechanical strength modulus (GPa) | 48 | 37 | higher strength → higher robustness ☺ | 38 |

[a] calculated from temperature-viscosity for F2 and Te glass, assuming the soft glass doping step for the F2/F2 fiber and the molten glass doping step for the Te fiber is both conducted at 650 °C. [b] calculated at wavelength of 700 nm.